\documentclass[12pt]{article}

\usepackage{amssymb}
\usepackage{amsfonts}
\usepackage{amsmath}
\usepackage{bm}
\usepackage{graphicx}
\usepackage{color}
\usepackage{cite}
\usepackage{graphicx}
\usepackage{wrapfig}
\usepackage[font=small,
labelfont=bf,
labelsep=period]{caption}

\definecolor{nv}{rgb}{0.1,0.1,0.6}
\definecolor{pr}{rgb}{0.2,0.1,0.5}
\definecolor{mg}{rgb}{0.4,0.0,0.4}

\newcommand{\nn}{\nonumber}

\newcommand{\beq}{\begin{equation}}
\newcommand{\eeq}{\end{equation}}
\newcommand{\beqy}{\begin{eqnarray}}
\newcommand{\eeqy}{\end{eqnarray}}
\newcommand{\beqyn}{\begin{eqnarray*}}
\newcommand{\eeqyn}{\end{eqnarray*}}

\newcommand{\bs}{\begin{slide}}
\newcommand{\es}{\end{slide}}
\newcommand{\bc}{\begin{center}}
\newcommand{\ec}{\end{center}}
\newcommand{\bmin}{\begin{minipage}}
\newcommand{\emin}{\end{minipage}}

\begin{document}

\begin{titlepage}
\begin{center}

{\Large\bf Comments on the Bl\"{u}mlein-B\"{o}ttcher\\
determination of higher twist corrections to\\[2mm]
the nucleon spin structure function $g_1$}

\end{center}
\vskip 2cm
\begin{center}
{\bf Elliot Leader}\\
{\it Imperial College London\\ Prince Consort Road, London SW7
2BW, England }
\vskip 0.5cm
{\bf Aleksander V. Sidorov}\\
{\it Bogoliubov Theoretical Laboratory\\
Joint Institute for Nuclear Research, 141980 Dubna, Russia }
\vskip 0.5cm
{\bf Dimiter B. Stamenov \\
{\it Institute for Nuclear Research and Nuclear Energy\\
Bulgarian Academy of Sciences\\
Blvd. Tsarigradsko Chaussee 72, Sofia 1784, Bulgaria }}
\end{center}

\vskip 0.3cm
\begin{abstract}
\hskip -5mm

In a recent analysis of the world data on polarized DIS, Bl\"{u}mlein
and B\"{o}ttcher conclude that there is no evidence for higher twist
contributions, in contrast to the claim of the LSS group, who find
evidence for significant higher twist effects. We explain the
origin of the apparent contradiction between these results.


\end{abstract}

\end{titlepage}

\newpage
\setcounter{page}{1}

\section{Introduction}
Because of the limited range of $Q^2$ available in the world data
sample for polarized DIS, and because of the great accuracy of
some of the data at relatively low values of $Q^2$, it is
important to be able to include the latter data in QCD analyses
aimed at extracting information on the polarized parton densities.
To do this consistently the theoretical formulae must be extended
beyond leading twist (LT) to allow for higher twist (HT)
contributions. This has been done for some years by Leader,
Sidorov and Stamenov (LSS) \cite{LSS_HT}, who have shown that such
contributions are important, especially in fitting the very
accurate CLAS data \cite{CLAS}. In contrast, Bl\"{u}mlein and
B\"{o}ttcher (BB) \cite{BB} claim that HT effects are completely
negligible. In
this note we explain that the disagreement is probably only
apparent, and results from attempting to compare two
incommensurate quantities.

\section{Structure of higher twist terms}

On the basis of the Operator Product Expansion (OPE), LSS have
utilized an expression for the experimental spin dependent
structure function $g_1(x,Q^2)_{exp}$  of the form
\beqy \label{LSS} g_1(x,Q^2)_{exp} &=& g_1(x,Q^2)_{LT}+
g_1(x,Q^2)_{TMC} + g_1(x,Q^2)_{HT} \nn \\
&=& g_1(x,Q^2)_{LT}+ g_1(x,Q^2)_{TMC} + \frac{h(x)}{Q^2}+{\cal O}(\Lambda^4/Q^4) , \eeqy
where the powers in $1/Q^2$ corrections to $g_1$ have been split
into the exactly known kinematical  target mass corrections (TMC)
and genuine dynamical higher twist terms (HT). In the LSS analysis
the logarithmic $Q^2$ dependence of $h$, which is unknown in QCD,
is neglected.
\begin{figure}[t]
\centering
\includegraphics[width=60mm]{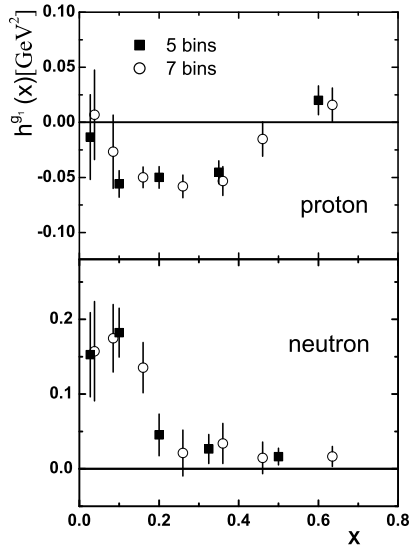}
\caption{Higher twist corrections extracted from the world
polarized DIS data (corresponding to 5 and 7 x bins). }
\label{HT_LSS06}
\end{figure}
Compared to the principal $1/Q^2$ dependence it is expected
to be small and the accuracy of the present data does not allow its determination. Therefore, the extracted from the data values of
$h(x)$ correspond to the mean $Q^2$ for each $x$-bin. The most recent results for $h(x)$ for proton and neutron targets \cite{LSS06} are
shown in Fig.1.

BB write the expression for $g_1(x,Q^2)_{exp}$ in a different
form, namely
\beq \label{BB}
g_1(x,Q^2)_{exp} = g_1(x,Q^2)_{LT}\big[ 1 +\frac{C(x)}{Q^2}\big], \eeq
where the assumption
\beq\label{assumption}
g_1(x,Q^2)_{HT} = g_1(x,Q^2)_{LT}\frac{C(x)}{Q^2} \eeq
is used for the twist-4 contribution, and any $Q^2$ dependence
in $C(x)$ is neglected.
(It is not clear whether the TMC are accounted for in $g_1(x,Q^2)_{LT}$.)

BB find no evidence for HT  i.e. their $C(x)$ for protons and
neutrons is compatible with zero. (There is a worrying issue
concerning their derivation of the neutron value, which we shall
comment on later.)

In trying to understand the apparent discrepancy with LSS,  BB
write: "This result is in disagreement to Ref. [21] (i.e. LSS \cite{LSS09}). Note that in the latter analysis a partonic
description of $F_1(x,Q^2)$ down to low values of $Q^2$ is used, while we
refer to the measured function."

This statement originates in the
fact that what is actually measured is effectively
$g_1(x,Q^2)_{exp}/F_1(x,Q^2)_{exp}$, so that $g_1(x,Q^2)_{exp}$
must be extracted from the data by multiplying by
$F_1(x,Q^2)_{exp}$.

However, the comment about Ref. [21]
(Ref. \cite{LSS09} in this paper) is totally incorrect. Indeed in \cite{LSS09} LSS write:

"According to this method, the $g_1/F_1$ and $A_1$  data have been
fitted using the experimental data for the unpolarized structure
function $F_1(x,Q^2)$
\beq
\left[\frac{g_1(x,Q^2)}{F_1(x,Q^2)}\right]_{exp}=\frac{g_1(x,Q^2)_{LT}
+g_1(x,Q^2)_{TMC} + h(x)/Q^2}{F_1(x,Q^2)_{exp}} .
\eeq
As usual, $F_1$ is replaced by its expression in terms of $F_2$ and
$R$, and phenomenological parametrizations of the experimental data for
$F_2(x,Q^2)$ [8] and the ratio $ R(x,Q^2)$ of the longitudinal to
transverse cross-sections [9] are used."

Hence the apparent discrepancy between BB and LSS is certainly not due to any difference in the handling of $F_1(x,Q^2)$.

\section{Why the discrepancy is only apparent}

On equating Eqs.~(\ref{LSS}) and (\ref{BB}) we have
\beq
C(x)=\frac{h(x)}{g_1(x,Q^2)_{LT} }~. \eeq

Thus, if it is legitimate
to neglect the scale dependence in $h(x)$ then $C(x)$ must vary
with $Q^2$, contradicting the use of $C(x)$ as
$Q^2$-independent. If, on the other hand, it is legitimate to
neglect the $Q^2$ dependence in $C(x)$, then  $h(x)$ must vary
with $Q^2$. We thus see that the two approaches are
incompatible and their results incommensurate. One of the two
methods (or perhaps both)
has to be incorrect and the fact that their results disagree is
inevitable and requires no further explanation. However, since the
LSS formulation is closer in structure to the  operator product
expansion, we believe it is more likely to be the correct way to
implement HT corrections. Moreover, the LSS results on HT are in
good agreement with those obtained from the study of the {\it
first} moments of the spin structure functions
$g_1^{(p,n)}(x,Q^2)$, and in particular, of the non-singlet
structure function $g_1^{(p-n)}$ (see Ref. \cite{n=1}).

One further point remains. BB utilize Eq.~(\ref{BB}) for proton
and deuteron data and then extract the neutron value of $C(x)$ via
\beq C_n(x) = \frac{2}{1- 1.5\omega_D}C_d(x) - C_p(x)  \eeq where
$\omega_D=0.05\pm 0.01$. From Eq.~(\ref{BB}) one sees that this is
incorrect. The correct relation should be \beq C_n(x) =
\frac{1}{g_{1n}(x,Q^2)_{LT}} \big[ \frac{2}{1-
1.5\omega_D}g_{1d}(x,Q^2)_{LT}C_d(x) - g_{1p}(x,Q^2)_{LT} C_p(x)
\big]  \eeq Thus, even if it is correct to take $C_p(x)$ and
$C_d(x)$ independent of $Q^2$, $C_n$ will inevitably inherit some
dependence on $Q^2$. Note also that the neutron spin structure
function $g_{1n}(x,Q^2)_{LT}$ passes through zero as a function of
$x$ and it is therefore dangerous to use the above equation to
extract the HT correction $C_n$.

\section{Conclusions}
We have shown that the LSS and BB methods of extracting HT
corrections in polarized DIS are incompatible and that it thus
makes no sense to compare their results - they are incommensurate.
We believe that the LSS approach, because it is closely related to
the operator product expansion, is more likely to be the correct
one.

\begin{center}
{\bf Acknowlwdgments}\\
\end{center}

This research was supported by the JINR-Bulgaria Collaborative
Grant, by the RFBR Grants (No 09-02-01149, 10-02-01259) and by the
Bulgarian National Science Foundation under Contract 288/2008. One
of the authors (E.L) is grateful to the participants at the
Brookhaven Summer Program on Nucleon Spin Physics for stimulating
comments, and to the organizers for support.

 \end{document}